\documentclass[aps,prl,reprint,showpacs,superscriptaddress,10pt]{revtex4-1}
\usepackage{amsmath,amssymb,amsfonts,graphicx,bm,longtable,dcolumn,color}
\usepackage{mathrsfs}
\usepackage[utf8]{inputenc}
\usepackage{textcomp}
\usepackage[T1]{fontenc}
\usepackage{lmodern}
\usepackage{dsfont}

\begin{document}

\title{Ballistic near-field heat transport in dense many-body systems}

\author{Ivan Latella}
\affiliation{Laboratoire Charles Fabry, UMR 8501, Institut d'Optique, CNRS, Universit\'{e} Paris-Saclay, 2 Avenue Augustin Fresnel, 91127 Palaiseau Cedex, France}

\author{Svend-Age Biehs}
\affiliation{Institut f\"{u}r Physik, Carl von Ossietzky Universit\"{a}t, D-26111 Oldenburg, Germany}

\author{Riccardo Messina}
\affiliation{Laboratoire Charles Coulomb (L2C), UMR 5221 CNRS-Universit\'{e} de Montpellier, F- 34095 Montpellier, France}
\affiliation{Department of Electrical Engineering, Princeton University, Princeton, NJ 08544, USA}

\author{Alejandro W. Rodriguez}
\affiliation{Department of Electrical Engineering, Princeton University, Princeton, NJ 08544, USA}

\author{Philippe Ben-Abdallah}
\email{pba@institutoptique.fr}
\affiliation{Laboratoire Charles Fabry, UMR 8501, Institut d'Optique, CNRS, Universit\'{e} Paris-Saclay, 2 Avenue Augustin Fresnel, 91127 Palaiseau Cedex, France}
\affiliation{Universit\'{e} de Sherbrooke, Department of Mechanical Engineering, Sherbrooke, PQ J1K 2R1, Canada.}

\begin{abstract}
  Radiative heat-transport mediated by near-field interactions is
  known to be superdiffusive in dilute, many-body systems. In this
  Letter we use a generalized Landauer theory of radiative heat
  transfer in many-body planar systems to demonstrate a nonmonotonic
  transition from superdiffusive to ballistic transport in dense
  systems. We show that such a transition is associated to a change of the polarization of dominant modes, leading to dramatically different
  thermal relaxation dynamics spanning over three orders of magnitude. 
  This result could have important consequences on thermal management at nanoscale of many-body systems.
\end{abstract}

\newcommand{\dif}{d}
\newcommand{\ee}{e}
\newcommand{\e}{e}
\newcommand{\ii}{i}
\newcommand{\iunit}{i}
\newcommand{\sub}[1]{#1}
\newcommand{\vect}[1]{\bm{#1}}
\newcommand{\kB}{k_B}

\maketitle

The theory of near-field radiative heat transfer has for many decades
remained largely limited to two-body
systems~\cite{PoldervH,LoomisPRB94,JoulainSurfSciRep05,VolokitinRMP07,Otey14,Song15}. Recently,
heat transport in many-body systems has also been considered in the
context of
nanoparticles~\cite{PBA-APL2006,PBA-PRB2008,PBA-PRL2011,Ordonnez} and
multilayer geometries, such as photonic
crystals~\cite{LauEtAl1,LauEtAl2} and hyperbolic
metamaterials~\cite{LiuNarimanov,BiehsEtAl2015,Messina-PRB2016}.  The
focus of much of this work has been the study of systems in which the
steady-state temperature distribution of a set of internal bodies is a
priori known and dictated via contact with large heat
reservoirs. There are, however, situations in which a full study of
heat transport necessitates an account of thermal relaxation through
radiative channels. A first step in this direction has been made by
generalizing Rytov's theory of fluctutational electrodynamics to
describe radiative transfer in many-body geometries with varying
temperature distributions, including nanoparticle
systems~\cite{Messina-2013,Yannopapas2013,Nikbakht2015}, multilayer
configurations~\cite{PBA-PRL2012,PBA-PRL2014-1,Messina-PRA2014,PBA-PRL2014-2,Dyakov,LatellaPRAppl15,PBA-PRL2016,Messina-PRB2016-1,Messina-PRB2016-2,Krugger},
and more generally, arbitrary geometries that include the possibility
of inhomogeneously varying temperature
profiles~\cite{Edalatpour16,Jin17}.  Furthermore, heat transport
within a collection of nanoparticles has been studied in
Ref.~\cite{BenAbdallahPRL13}, and the existence of a superdiffusive
regime has been found, albeit within a dipolar approximation.


In the present Letter, we employ a recently developed, exact
theoretical framework~\cite{LatellaPRB17} to investigate near-field
radiative heat transport in $N$-body systems consisting of parallel
planar slabs separated by vacuum, in which radiation is the only
source of thermal relaxation. We show that the temperature dynamics
and steady-state profile of the system depend strongly on geometric
parameters such as the system density, which imply different
heat-transport regimes. In particular, we prove the existence of a
nonmonotonic transition between a superdiffusive regime, previously
observed in Ref.~\cite{BenAbdallahPRL13}, and a ballistic regime that
appears in denser media and that also leads to dramatically faster
relaxation dynamics.  We also show that this transition is associated
with a change in the polarization of the dominant modes in the
transport.  In contrast to heat exchange in two-body geometries, where
near-field heat transfer is dominated by transverse-magnetic (TM)
modes, we found that transport in dense, many-body systems can have a
significant contribution from transverse-electric (TE) modes.


\begin{figure}[t!]
\includegraphics[width=0.9\columnwidth]{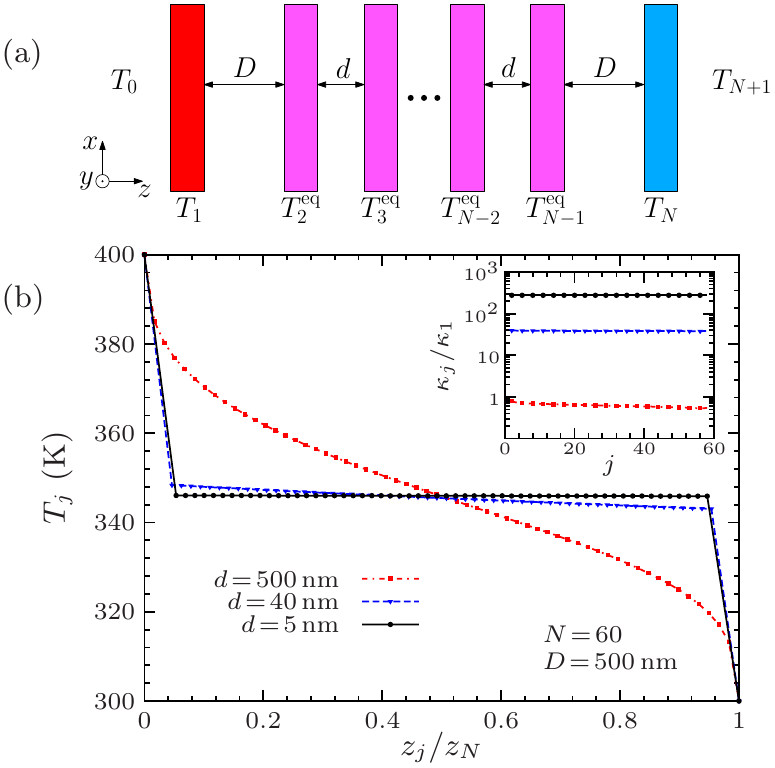}
\caption{(a) Schematic of an $N$-body system comprising $N-2$ planar
  slabs (purple) interacting with one another and with two external
  slabs at fixed temperatures $T_1$ (red) and $T_N$ (blue). All
  internal separation distances $d$ are identical while the coupling
  to the external thermostats depends on the separation distance
  $D$. In the steady state, each internal slab reaches a local
  equilibrium temperature $T_j^\mathrm{eq}$. (b) Steady-state temperature
  profile as a function of the normalized position $z_j/z_N$ for a
  system of $N=60$ SiC slabs of thickness 200\,nm, for different $d$
  and fixed $D=500\,$nm. The inset shows the ratio of the effective
  internal conductivities $\kappa_j/\kappa_1$ (see text).}
\label{fig1}
\end{figure}

\begin{figure*}[t!]
\includegraphics[width=1.8\columnwidth]{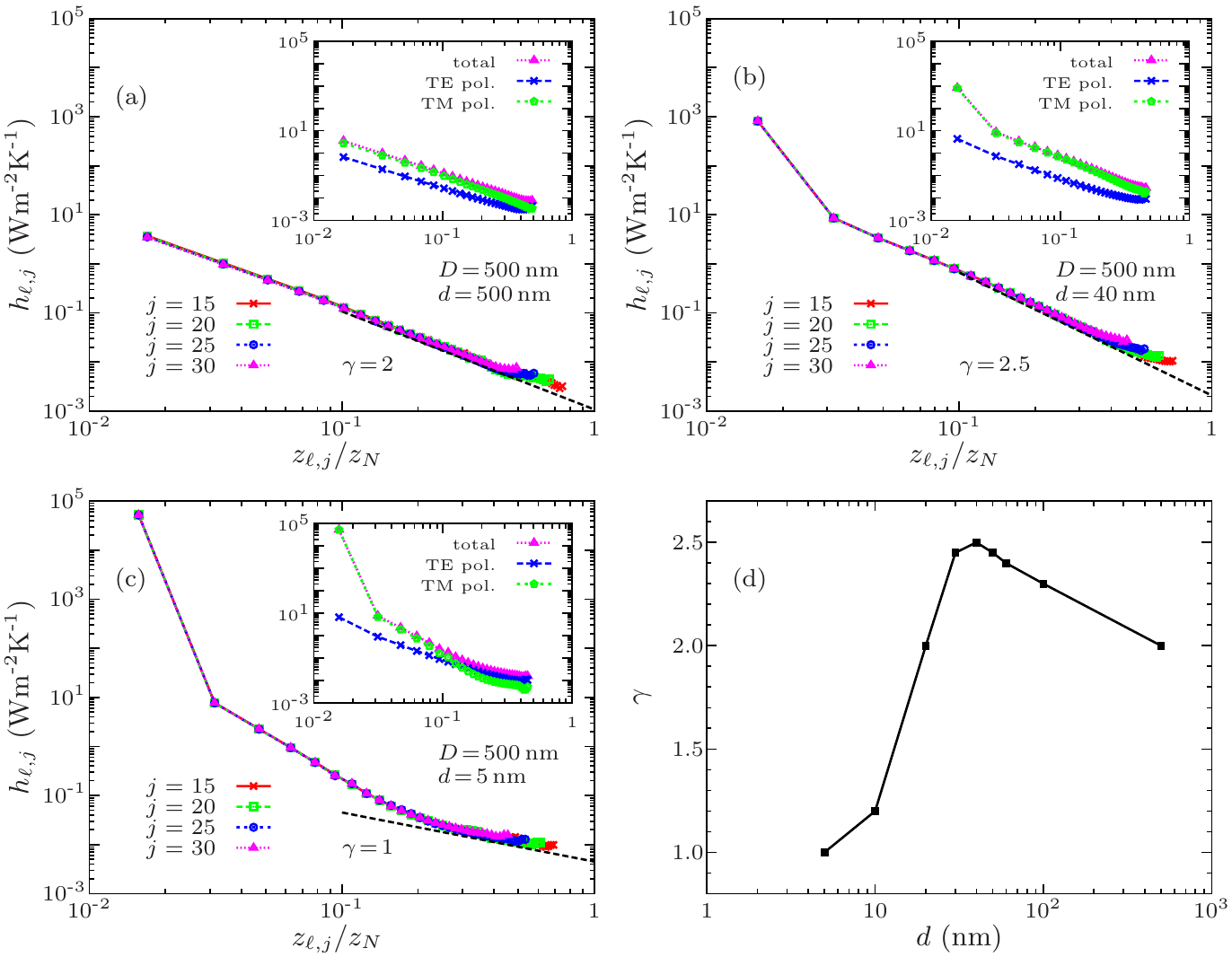}
\caption{Heat-transfer coefficients $h_{\ell,j}$ (see text) with
  respect to the normalized separation $z_{\ell,j}/z_N$, for fixed
  values of $j$ and $D=500\,$nm, and three values of (a)
  $d=500\,$nm, (b) $d=40\,$nm and (c) $d=5\,$nm. Dashed lines indicate
  the asymptotic behavior of $h_{\ell,j} \sim 1/z_{\ell,j}^\gamma$
  at large separations. The value of $\gamma$ indicates the nature of
  the heat-transport regime, from superdiffusive ($1<\gamma<3$) to
  ballistic ($\gamma\to 1$). The insets decompose $h_{\ell,j}$ for
  $j=30$ into TE and TM polarization contributions. (d) Exponent
  $\gamma$ as a function of $d$.}
\label{fig2}
\end{figure*}

To begin with, let us consider a system composed of $N$ planar slabs
separated by vacuum, orthogonal to the $z$ axis and assumed to be
infinite in the $x$ and $y$ directions, as sketched in
Fig.~\ref{fig1}(a). The thicknesses $\delta_j$ of the bodies are
assumed to be equal, $\delta_j=\delta$ for $j=1,\dots,N$, and below we
take $\delta=200\,$nm. The temperatures of slabs 1 and $N$, referred
to as \emph{external} slabs, are held constant at $T_1=400\,$K and
$T_N=300\,$K, respectively, via contact with an external reservoir,
while all the other \emph{internal} slabs are allowed to reach their
own equilibrium temperature $T_j^\mathrm{eq}$ ($j=2,...,N-1$). We also
consider that the system is immersed in an environment (thermal bath)
at temperature $T_0=T_{N+1}=T_B=300\,$K. Below $D$ denotes the
distance between slabs 1 and 2, as well as slabs $N-1$ and $N$,
whereas $d$ is the distance between adjacent, internal
slabs. Furthermore, in our numerical simulations we assume that all
the bodies are made of silicon carbide (SiC), whose permittivity at
frequency $\omega$ can be described by the Drude-Lorentz
model~\cite{Palik} $\varepsilon(\omega)=\varepsilon_\infty
\frac{\omega^2_L-\omega^2-i\Gamma\omega}{\omega^2_T-\omega^2-i\Gamma\omega}$,
where $\varepsilon_\infty=6.7$ is the high frequency dielectric
constant, $\omega_L=1.83\times 10^{14}\,$rad/s is the longitudinal
optical phonon frequency, $\omega_T=1.49\times 10^{14}\,$rad/s is the
transverse optical phonon frequency, and $\Gamma=8.97\times
10^{11}\,$rad/s is the damping rate. As shown in
Ref.~\cite{LatellaPRB17}, the net radiative flux per unit surface
received by any given slab $j$ can be written as a sum over the energy
exchanged with every other body $\varphi_{\ell,j}$, with
\begin{equation}
\varphi_j=\sum_{\ell\neq j}\varphi_{\ell,j}=\sum_{\ell\neq j}\int_0^\infty\!\!\frac{d\omega}{2\pi}\int_0^\infty\!\!\frac{dk}{2\pi}k \sum_{p}\hbar\omega\,n_{\ell,j} \mathcal{T}^{\ell,j},
\label{energy_flux}
\end{equation}
where $\ell \neq j$ runs from 0 to $N+1$ (including the
external environment). In this expression $p=\mathrm{TE},\mathrm{TM}$
denotes the two polarizations, $k$ is the parallel component of the
wave vector, and $n_{\ell,j}\equiv n_\ell - n_j$, with
$n_j=\left(\ee^{\hbar\omega/\kB T_j }-1\right)^{-1}$ denoting the
Bose distribution. The Landauer coefficient
$\mathcal{T}^{\ell,j}=\mathcal{T}^{\ell,j}(\omega,k,p)$, which can
vary between 0 and 1, describes the contribution of each mode
$(\omega,k,p)$ to the energy exchange and depends on the geometrical
and material properties of the slabs~\cite{LatellaPRB17}. The local
equilibrium temperatures $T_j^\mathrm{eq}$ of the internal slabs can
be calculated by requiring that in the steady state, the net flux
received by each slab is zero, that is by solving the system of
transcendental equations, $\varphi_j = 0$ for $j=2,\dots,N-1$. The
steady-state temperature profiles inside the system are shown in
Fig.~\ref{fig1}(b) for $N=60$ slabs and for several separation
distances $d\in\{5,40,500\}\,$nm and fixed $D=500\,$nm. We first
observe that, while for $d=D=500\,{\rm nm}$ the temperature profile
decays smoothly, the configurations having a smaller $d$ reveal a more
dramatic jump between the external and the adjacent $(T_2,T_{N-1})$
temperatures, in which case the internal slabs become much more
thermally isolated from the reservoirs. Moreover, the shape of the
profile clearly depends on $d$, becoming nonlinear for $d=500\,$nm,
close to linear for $d=40\,$nm, and nearly constant for $d=5\,$nm.

We now describe how the main features characterizing heat transport in
this geometry, i.e. the temperature profile near the boundary and
within the bulk, depend on both $D$ and $d$. As far as the former is
concerned, the main parameter of interest is the relative coupling
strength of boundary versus internal slabs, quantified by defining an
effective, thermal conductivity
$\kappa_j=\varphi_{j,j+1}d_j/(T_j-T_{j+1})$, where $d_1=d_{N-1}=D$ and
$d_j=d$ for $j=2,\dots,N-2$.  The ratio $\kappa_j/\kappa_1$, which can
be interpreted as a measure of the boundary thermal resistance, is
plotted in the inset of Fig.~\ref{fig1}(b), showing that $\kappa_j$ is
almost constant within the chain of internal slabs and that
$\kappa_j/\kappa_1$ is close to unity for $d=500\,$nm, increases with
decreasing $d$, and reaches two orders of magnitude when $d=5\,$nm. As
illustrated in Fig.~\ref{fig3} below, the smoothness of the
temperature profile near the boundary only depends on the ratio
$d/D$. We next focus on the shape of the temperature profile within
the bulk, which is closely related to the transport regime and
requires a more nuanced description of the problem.


As show in Ref.~\cite{BenAbdallahPRL13}, to understand and classify
the various transport regimes in this kind of system, it is useful to
study the power exchanged between layers in the limit of large
$N$. For convenience, we make the simplifying assumption that the
temperature differences involved in the system are small enough to
allow a linearization of $n_{\ell,j}$. Under this assumption, the net
flux on slab $j$ reads,
\begin{equation}
  \varphi_j\simeq\sum_{\ell \neq j} h_{\ell,j}(T_\ell-T_j),
\label{Eq:power_lin}
\end{equation}
where we have introduced the heat-transfer coefficients,
\begin{equation}
h_{\ell,j}=\int_0^\infty\!\!\frac{d\omega}{2\pi}\int_0^\infty\!\!\frac{dk}{2\pi}k \sum_{p} \hbar\omega\frac{\partial n_{j}}{\partial T_j} \mathcal{T}^{\ell,j}.
\label{heat_transfer_coeff}
\end{equation}
Assuming that $h_{\ell,j} \sim 1/z_{\ell,j}^\gamma$, for some exponent
$\gamma = 1+\alpha$, where $z_{\ell,j} \equiv |z_\ell-z_j|$ and $z_j$
denotes the position of the $j$-th layer, one finds that in the
continuum limit of $N \to \infty$ (with fixed $Nd$), $T_j\to T(z)$ and
\begin{equation}
  \varphi_{j} \to \varphi(z)\sim(-\Delta)^{\alpha/2}T(z),\label{Eq:power_2}
\end{equation}
where $(-\Delta)^{\alpha/2}$ is the fractional
Laplacian defined in 1D systems as
($0<\alpha<2$)
\begin{equation}
(-\Delta)^{\alpha/2}T(z)=c_{\alpha}\,\text{PV}\int_{-\infty}^\infty\frac{T(z)-T({z'})}{|z-{z'}|^{1+\alpha}}d{z'}\label{frac},
\end{equation}
with $c_{\alpha}$ a constant~\cite{Podlubny,Samko} and where $\text{PV}$ denotes the principal value. As we
discuss below, Eq.~\eqref{frac} can be used as a tool to relate the
asymptotic, large-distance behavior of $h_{\ell,j}$ to the regime of
heat transport. It follows from Eq.~\eqref{frac} that the regime of
heat transport is superdiffusive when $1 < \gamma < 3$. In the
limiting case $\gamma \to 3$, the fractional Laplacian degenerates
into its classical form and the regime of transport is diffusive. On
the other hand, as $\gamma \to 1$, the fractional Laplacian approaches
the identity operator and the transport becomes ballistic.

\begin{figure}
\includegraphics[width=0.9\columnwidth]{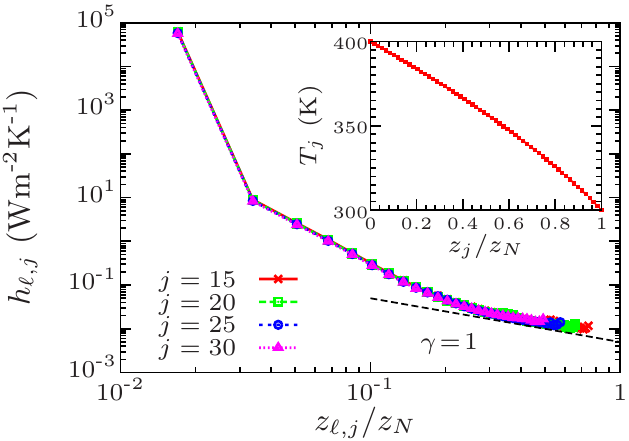}
\caption{Heat-transfer coefficients $h_{\ell,j}$ as a function of the
  normalized separation $z_{\ell,j}/z_N$, for $D=d=5\,$nm, along with
  the corresponding steady-state temperature profile (inset).} 
\label{fig3}
\end{figure}

Figure~\ref{fig2} shows $h_{\ell,j}$ for multiple values of $j$ as a
function of $\ell$, for the same system of Fig.~\ref{fig1}. When
$d=500\,$nm, corresponding to a SiC volume fraction of 28.5\% (dilute
system), $h_{\ell,j}$ asymptotically decays as $1/z_{\ell,j}^{2}$,
showing that indeed the heat transport is superdiffusive, as in simple
dipolar systems~\cite{BenAbdallahPRL13}. Note that the small
variations in $h_{\ell,j}$ at the extreme end of the curves come from
finite-size effects and are therefore not taken into account in the
scaling analysis. When $d=100\,$nm [see Fig.~\ref{fig2}(b)], the
exponent in the scaling of $h_{\ell,j}$ increases, but the transport
regime remains superdiffusive. On the other hand, when $d=5\,$nm,
corresponding to a SiC volume fraction of 97.5\% (dense system),
$h_{\ell,j} \sim 1/z_{\ell,j}$, in which case the transport is
ballistic and the system experiences an effectively weak thermal
resistance within the bulk. Figure~\ref{fig2}(d) shows $\gamma$ as a
function of $d$, illustrating a nonmonotonic behavior as the system
transitions from a superdiffusive to a ballistic regime. Furthermore,
as illustrated on the insets of Fig.~\ref{fig2}, which show the
contributions of TE and TM modes to $h_{\ell,j}$, we find that TE
modes dominate and hence determine the (ballistic) transport regime at
small $d$; in contrast, TM modes are the main heat carriers in the
superdiffusive regime, which is the case in typical two-body
geometries involving polaritonic resonances. This surprising result is
a clear indication of the complexities and richness of heat transport
in many-body systems.

Figure~\ref{fig3} shows $h_{\ell,j}$ along with the temperature
profile (inset) for $d = D = 5\,{\rm nm}$. Comparing the former to the
results in Fig.~\ref{fig2}(c), one confirms that the transport regime
is independent of $D$ and therefore only depends on the density within
the bulk, determined by $d$. On the other hand, comparing the
temperature profile in Fig.~\ref{fig3} to those in Fig.~\ref{fig1},
one infers that indeed only the ratio $d/D$ (or thermal resistance)
controls the smoothness of the profile near the boundaries.

\begin{figure}
\includegraphics[width=\columnwidth]{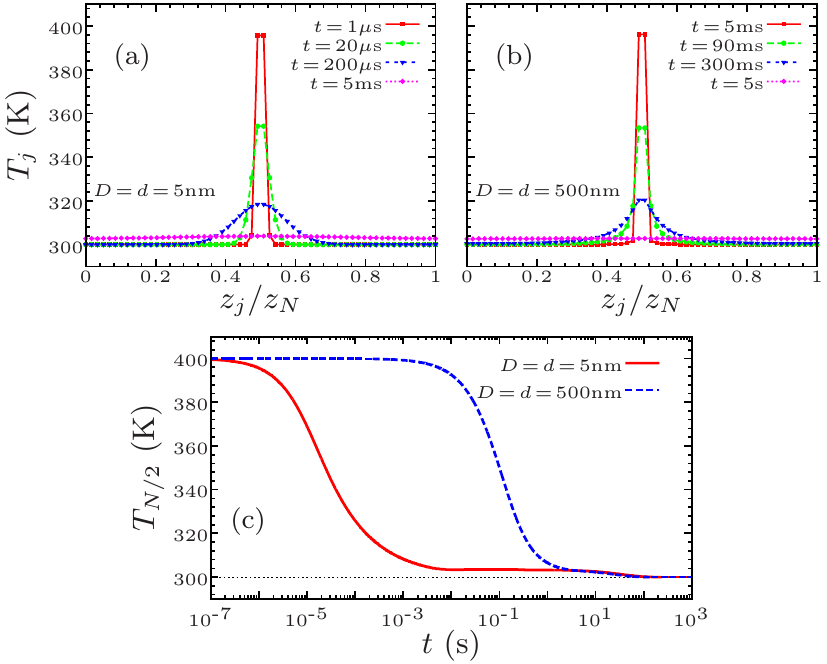}
\caption{Temporal evolution of the temperature profile for a system
  ($N=60$) interacting with a thermal bath at
  $T_B=T_0=T_{N+1}=300\,$K. At $t=0$, all bodies have temperature
  $T_B$ except the two central slabs, which have
  $T_{N/2}=T_{N/2+1}=400\,$K. (a) Ballistic regime ($D=d=5\,$nm). (b)
  Superdiffusive regime ($D=d=500\,$nm). (c) Temperature of slab $N/2$
  as a function of time in the two previous cases.}
\label{fig4}
\end{figure}

Finally, we investigate the impact of the previously considered transport regimes on
the relaxation dynamics of the system.  Given an initial temperature
distribution $\vect{T}(0) = \bigl(T_1(0),...,T_N(0)\bigr)$, the
temperatures of the bodies $\vect{T}(t) = \bigl(T_1(t), \dots
,T_N(t)\bigr)$ at any give time $t > 0$ are solutions of the energy
balance equation,
\begin{equation}
\partial_t \vect{T} = \mathds{K} \cdot \vect{T} + \vect{S}, 
\label{Eq:diff}
\end{equation}
where $\mathds{K} = \mathds{H}/(C \delta)$ is a stiffness matrix
defined in terms of the heat-transfer matrix $\mathds{H}$, with
elements $[\mathds{H}]_{\ell,j}=h_{\ell,j}$ ($\ell, j = 1,\dots,N$),
and the SiC heat capacity per unit volume~\cite{SommersATE10} $C =
8.15\,$J$\cdot $cm$^{-3}\cdot$K$^{-1}$. Here, $h_{j,j} = -\sum_{\ell
  \neq j} h_{\ell,j} $ quantifies the emission rate of body $j$ in the
presence of the other slabs, while $\vect{S}=\frac{T_B}{C \delta}(
h_{0,1}+h_{N+1,1}, \dots , h_{0,N} + h_{N+1,N})$ denotes the source
term corresponding to power supplied by the bath to each
layer. Equation~(\ref{Eq:diff}) is simply a discrete form of the
fractional diffusion equation, the fractional exponent being related
to the scaling of $h_{\ell,j}$, whose solution in the steady state
reads $\vect{T}^\mathrm{eq} = -\mathds{K}^{-1} \cdot \vect{S}=(T_B,
\dots ,T_B)$. Since $h_{\ell,j}$ depends weakly on $T_j$, we assume
that $\mathds{K}$ is a time-independent matrix, in which case the time
evolution of the temperature profile is thus given by $\vect{T}(t) =
\exp(\mathds{K} t) \cdot [ \vect{T}(0) - \vect{T}^\mathrm{eq} ] +
\vect{T}^\mathrm{eq} $. Figure~\ref{fig4} shows the temporal evolution
of the system in both superdiffusive and ballistic regimes, assuming
an initial temperature profile corresponding to heating of the two
central slabs to a temperature of 400\,K. We observe a strong increase
of the relaxation dynamics in the ballistic regime compared to the
superdiffusive case, showing a difference in characteristic
equilibration scales of nearly three orders of magnitude (from
microseconds to milliseconds for a reduction of about half the initial
overheating).  We also observe that, as previously observed in dilute
media~\cite{Messina-2013}, the relaxation process occurs in two
distinct timescales. First, all layers thermalize at the same
temperature through near-field interactions in about $5\,$ms in dense
media (seconds in the diluted case). Subsquently, all layers
collectively cool down to the ambient temperature through far-field
interactions with the thermal bath.

In conclusion, we have studied a many-body geometry of planar slabs
which exhibits a transition in the regime of radiative heat-transport,
from ballistic to superdiffusive, with respect to slab density. Our
predictions reveal complex, many-body effects in addition to
dramatically different relaxation dynamics, depending on the transport
regime. These effects could have important implications for
thermal management at nanoscale in devices involving multiple, interacting
elements thermally coupled in the near field.

\emph{Acknowledgments}---We acknowledge the use of the computing
center M\'{e}soLUM of the LUMAT research federation (FR LUMAT
2764). This work was partially supported by the National Science
Foundation under Grant no. DMR-1454836 and by the Princeton Center for
Complex Materials, a MRSEC supported by NSF Grant DMR-1420541.

\end{document}